\documentclass[12pt,a4paper]{article}
\usepackage{mathrsfs}
\usepackage{epsfig}

\usepackage{tabularx}
\usepackage{graphics}
\usepackage{graphicx}
\usepackage{float}
\usepackage{multirow}
\usepackage{tabularx}

\begin{document}
\title{Sterile neutrino and leptonic decays of the pseudoscalar mesons }
\author{Chong-Xing Yue and Ji-Ping Chu   \\
{\small Department of Physics, Liaoning  Normal University, Dalian
116029, P. R. China}
\thanks{E-mail:cxyue@lnnu.edu.cn}}
\date{\today}

\maketitle
\begin{abstract}

 \vspace{0.5cm} We consider a scenario with only one sterile neutrino $ N $, negligible mixing with the active neutrinos $ \nu_{L} $, where its interactions with ordinary particles could be described in a model independent approach based on an effective theory. Under such a framwork, we consider the contributions of the sterile neutrino $N$ to the pure leptonic decays $ M \rightarrow l\nu $ and $ \nu \overline{\nu} $, and the radiative leptonic decays $ M \rightarrow l\nu\gamma $ with $M$ denoting the pseudoscalar mesons $ B $, $ D $ and $ K$. We find that it can  produce significant effects on some of these rare decay processes.

\end{abstract}
\newpage

\noindent{\bf 1. Introduction }\vspace{0.5cm}

Discovery of the 125 GeV scalar particle seeming to be the Higgs boson predicted by the standard model (SM) [1, 2] is the crowning achievement of the LHC Run I. Although most of the experimental measurements are in good agreement with the SM predictions, there are still some unexplained discrepancies and theoretical issues that the SM can not solve. Furthermore, experiments in the last decades have confirmed that at least two of the three known neutrinos must have nonzero masses and lepton flavors are mixed [3], which is so far the most clear experimental evidence for the existence of new physics beyond the SM.

In order to naturally explain the tiny neutrino masses, sterile neutrinos are usually introduced in most of the new physics scenarios. The seesaw mechanism is one of the simple paradigms for generating suitable neutrino masses [4]. Different realizations of this mechanism give rise to sterile neutrinos with mass covering various mass ranges. So it is reasonable to search for the direct and indirect signals of the sterile neutrinos with masses in the broad range from eV to TeV.

If the sterile neutrino mass is below the electroweak scale (i.e. $ m_N < m_W$), it can behave as long-lived particle with a measurable decay length, which give us an opportunity to probe its signatures by taking advantage of the displaced vertex techniques. So far there are many studies on searches for long-lived sterile neutrinos in the LHC and future colliders (see for instance [5, 6] and references therein). Furthermore, if the sterile neutrinos are sufficiently light, they may induce important impact on electroweak precision and many other observables. For example, they might be produced via heavy meson decays and further contribute to the rare meson decay processes (see for instance [7, 8, 9] and references therein). In this paper we will consider the effects of the sterile neutrino on the pure leptonic decays $ M \rightarrow l\nu $ and $ \nu \overline{\nu} $ with $ M $ denoting the pseudoscalar mesons $B$, $D$ and $K$, and the radiative leptonic decays $ M \rightarrow l \nu \gamma $  in the context of a general effective theory framework.

The sterile neutrino interactions can be described in a model independent approach based on an effective theory [10]. We assume that the interactions of the sterile neutrino $N$ with ordinary particles arise from high dimension effective operators and are dominant in comparison with the mixing with light neutrinos through the Yukawa couplings. The different operators  in the effective Lagrangian parameterize a wide variety of UV-compete  new physics models. Thus, considering their possible contributions to specific physical processes can give us smoking gun on what kind of new physics at higher energy range is responsible for the observables. The relevant phenomenological researches have been addressed in recent works [11, 12, 13]. The main goal of this paper is to consider a most simple scenario with only  one sterile neutrino $N$, which has a negligible mixing with the SM neutrinos $\nu_{L}$ and interacts with ordinary particles by effective operators of higher dimension, and see whether $ N $ can produce significant contributions to the decay processes $ M \rightarrow l\nu $, $ l\nu\gamma $ and $\nu \overline{\nu} $.

This paper is organized as follows. In section 2, we first review the relevant effective operators and the existing constraints on the effective coupling constants, and then calculate the contributions of the  sterile neutrino $N$ to the decay processes $ M \rightarrow l\nu $, $l\nu\gamma $ where $M $ denotes the pseudoscalar meson $B$, $D$ or $K$, $ l $ and $ \nu$ represent the SM charged leptons and neutrinos, respectively. Its effects on the decay processes $ M \rightarrow \nu \overline{\nu} $ are studied in section 3. Our conclusions are given in section 4.

 \vspace{0.5cm} \noindent{\bf 2. The sterile neutrino  $ N $ and the decay processes $ M\rightarrow l\nu $ and $ l\nu\gamma $
 }

  \vspace{0.5cm} \noindent{\bf 2.1 The relevant effective couplings of the sterile neutrino $N$  }

\vspace{0.5cm}The effects of the new physics involving one sterile neutrino and the SM fields can be parameterized by a set of effective operators $ O_{J} $ satisfying the $ SU(2)_{L} \times U(1)_{Y} $ gauge symmetry [14]. The contributions of these operators to observables are suppressed by inverse powers of the new physics scala $ \Lambda $. The total Lagrangian is written as
\begin{eqnarray}
\mathcal{L}&= &\mathcal{L}_{SM} + \sum_{n=5}^{\infty} \frac{1}{\Lambda^{n-4}} \sum_{J} \alpha_{J} O_{J}^{n},
\end{eqnarray}
where $ O_{J}^{n} $ are gauge-invariant local operators with mass dimension $ n $.

In the case of neglecting the sterile-active neutrino mixing, the dimension 5 operators do not contribute to the studied decay processes, we will only consider the contributions of the dimension 6 operators. The decay processes considered in this paper only involve a meson, we can neglect all operators with a tensor Lorentz structure, because of their vanishing hadronic matrix element. Then the effective Lagrangian derived from the relevant operators, which produce main contributions to the decay processes $ M \rightarrow l\nu$, $l\nu\gamma $ and $ M \rightarrow \nu \overline{\nu} $, can be written as
\begin{eqnarray}\mathcal{L}_{eff}&= &\frac{1}{\Lambda^{2}}\{ \frac{m_{Z} v }{2}\alpha_{Z} Z^{\mu}\overline{N_{R}} \gamma_{\mu} N_{R} - \frac{m_{W} v }{\sqrt{2}}\alpha_{W}^{i} W^{+\mu} \overline{N_{R}} \gamma_{\mu} l_R^i  \\
&&-i\sqrt{2} v( c_W \alpha_{NB}^{i}+  s_W \alpha_{NW}^{i})(P^{A}_{\mu}~ \overline{ \nu_{L}^{i}} \sigma^{\mu\nu}N_R~ A_{\nu}) \nonumber \\ 
&&\left.+ \alpha_{V_{0}}^i \overline{d_R^i} \gamma^{\mu} u_{R}^{i} \overline{N_{R}} \gamma_{\mu} l_{R}^{i} + \alpha_{V_{3}}^{i} \overline{u_{R}^{i}} \gamma^{\mu} u_{R}^{i} \overline{N_{R}} \gamma_{\mu} N_{R} + \alpha_{V_{4}}^{i} \overline{d_{R}^{i}} \gamma^{\mu} d_{R_{i}} \overline{N_{R}} \gamma_{\mu} N_{R} + h.c.\}. \right.\nonumber
\end{eqnarray}
Where $ \alpha's$ are the effective coupling constants, $ v \approx 246 $ GeV is the electroweak symmetry breaking scale and a sum over the family index $i$ is understood. $ s_{W}= \sin\theta_{W} $ and $ c_{W}= \cos\theta_{W} $ with $ \theta_{W} $ being the Weinberg angle, $-P^{A}$ is the 4-momentum of the outgoing photon. Considering the one-loop coupling constants are naturally suppressed by a factor $ 1 / 16 \pi^{2} $ [10, 15], in above equation, we have not shown the terms generated by one-loop operators, because their contributions to the meson decays considered in this paper are much smaller than those of the tree-level operators. However, the sterile neutrino $N$ might generate significant  contributions to the decay  $ M \rightarrow l \nu \gamma $ via the process $ M \rightarrow lN \rightarrow l \nu \gamma $. Thus we give the relevant effective Lagrangian terms related the decay $N \rightarrow \nu \gamma $, although they are induced  by one-loop level tensorial operators:
\begin{eqnarray}
O_{NB} = (\overline{L}\sigma^{\mu\nu}N)\widetilde{\phi}B_{\mu\nu},~~~~~O_{NW} = (\overline{L}\sigma^{\mu\nu}\tau^{I}N)\widetilde{\phi}W^{I}_{\mu\nu}.
\end{eqnarray}
The first and second terms of Eq.(2) are associated to the following operators:
\begin{eqnarray}
O_{NN\phi} = i(\phi^{+}D_{\mu}\phi)(\overline{N}\gamma^{\mu}N),~~~~~O_{Nl\phi} = i(\phi^{T}\epsilon D_{\mu}\phi)(\overline{N}\gamma^{\mu}l_{i}).
\end{eqnarray}
Where $B_{\mu\nu}$ and $W^{I}_{\mu\nu}$ denote  the $U(1)_Y$ and $SU(2)_L$ field strengths, respectively. $\gamma^{\mu}$ and $\sigma^{\mu\nu} $ are the Dirac matrices, $\epsilon = i\sigma^{2} $ is the antisymmetric symbol. Taking the scalar doublet  as $\phi= \left(\begin{array}{cc}
0\\
(v+h)/\sqrt{2}
\end{array}
\right)  $ with $h$ being the Higgs field, after spontaneous symmetry breaking of the SM gauge group, the above operators can give Eq.(3) and the terms of Eq.(2) involving the electroweak gauge bosons $W$ and $Z$.

It is well known that the sterile-active neutrino mixing parameters in various seesaw models are severely contained by electroweak precision measurement data and direct collider searches. Ref. [13] has translated these existing bounds into the constraints on  $ \alpha's $. They have shown that the most stringent constraints on the couplings involving the first generation fermions come from the $ 0 \nu \beta \beta $ decay and there is $ \alpha_{0 \nu \beta \beta} \leq 3.2 \times 10^{-2} \sqrt{(m_{N}/100)}$ for $ \Lambda=1 $ TeV, while the other ones should satisfy $ \alpha \leq 0.32 $ given by the electroweak precision data. In our following numerical estimation, we will consider these constraints and take their maximal values.

\vspace{0.5cm} \noindent{\bf 2.2 The pseudoscalar meson decays $ M \rightarrow l\nu $   }

\vspace{0.5cm}In the SM, the leading order amplitude for the $ M^{-} \rightarrow l \overline{\nu_{l}} $ decay is

\begin{eqnarray}A_{SM}=\frac{4 G_{F} V_{ij}}{\sqrt{2}} \left \langle {l \overline{\nu_{l}}}  \left | \overline{l_{L}} \gamma^{\mu} \nu_{L} \right | {0} \right \rangle \left \langle {0} \left |\overline{u_{L}^{i}} \gamma_{\mu} d_{L}^{j} \right | {M^{-}} \right \rangle,
\end{eqnarray}
where $ i$, $j $ are the quark flavor indices of the corresponding pseudoscalar meson and $ V_{ij} $ is the CKM matrix element. Defining the meson decay constant, $ \left \langle { 0 } \left | \overline{u^{i}} \gamma^{\mu} \gamma_5 d^{i} \right | {M^{-}(P)} \right \rangle = i F_{M} P^{\mu}  \ $, then the decay width can be written as [16]

\begin{eqnarray} \Gamma(M^{-} \rightarrow l \overline{\nu_{l}}) =\frac{G_{F}^{2}}{8\pi} \left | V_{ij} \right |^{2} F_{M}^{2} m_{l}^{2} m_{M} (1-\frac{m_{l}^{2}}{m_{M}^{2}})^{2}. \end{eqnarray}

If we include the electroweak and radiative corrections [17], the decay width should be written as $ \Gamma^{'}(M \rightarrow l\nu)=(1+\sigma) \Gamma(M \rightarrow l\nu) $. However, in this paper, we will focus our attention on the relative correction effects of the sterile neutrino $N$ on the decay $ M \rightarrow l\nu $, so we will do not include the correction contributions in our numerical estimation. The numerical values of the pseudoscalar meson mass $ m_{M} $, the corresponding decay constant $ F_{M} $, $ V_{ij} $, and the lepton mass $ m_{l} $ used in our numerical calculations are taken from Ref.[3].

 The decay process $ M \rightarrow l\nu $ is helicity suppressed in the SM, which is sensitive to new physics effects (for example see Ref. [18]), and thus is of great interest as a probe for new physics beyond the SM. If the sterile neutrino $N$ is sufficiently light, i.e. $ m_N < m_M $, it can be on-shell produced from meson decays. If its decay length, which can be obtained from its total decay width, is larger than the size of the detector, then it does not decay in the detector and appears as missing energy, which is similar to the active neutrino. In this case, the sterile neutrino $N$ can change the branching ratio $ Br( M \rightarrow l\nu) $. In the scenario considered in this paper, using the relevant couplings given by Eq.(2), we can obtain the expression form for the decay width $ \Gamma ( M^{-} \rightarrow l\overline{N}) $
\begin{eqnarray}&&\Gamma (M^{-} \rightarrow l\overline{N})
 \\
&&= \frac{m_{M}^{3} F_{M}^{2}} {8 \pi} \left | V_{ij} \right |^{2}
 [G_{F}^{2} (\frac{\alpha_{W}^{l} v^{2}}{2 \Lambda^{2}})^{2}+(\frac{\alpha_{V_{0}}^{l}}{\Lambda^{2}})^{2}]  \lambda^\frac{1}{2}(1,y_{l},y_{N}) [y_{l}+y_{N}-y_{l}^{2}-y_{N}^{2}+2 y_{l} y_{N}] \nonumber
 \end{eqnarray}
with $ \lambda(a,b,c)=(a^{2}+b^{2}+c^{2} - 2 a b - 2 a c - 2 b c)$, $ y_{l}=m_{l}^{2} / m_{M}^{2} $ and $ y_{N}= m_{N}^{2} / m_{M}^{2} $. It is obvious that the decay width $ \Gamma ( M^{+} \rightarrow l^{+} N) $ has similar fashion.

 In order to analyse the relative strength of the SM and the sterile contributions, we define the ratio
\begin{eqnarray} R=\frac{\Gamma^{SM}(M \rightarrow l \nu_{l})+\Gamma (M \rightarrow l N)}{\Gamma^{SM}(M \rightarrow l \nu_{l})}.
\end{eqnarray}
In above equation, we have ignored the interference effects, because the interference term between two kinds of contributions being proportional to the factor $ m_{\nu}m_{N} $ with $ m_{\nu} \approx 0 $ [9]. Our numerical results about the positive charged pseudoscalar mesons are summarized in Table 1, which are obtained in the case of $ \Lambda=1 $ TeV,  $ \alpha_{W}^{e}\simeq  \alpha_{V_{0}}^{e}= 3.2 \times 10^{-2} \sqrt{(m_{N}/100)}$ and $ \alpha_{W}^{l}\simeq  \alpha_{V_{0}}^{l}=  0.32 $ with $l=\mu$ and $\tau$. In Table 1, we also show the values of the parameter $R^{exp} = Br^{exp}(M^{+} \rightarrow l^{+} \nu )/ Br^{SM}(M^{+} \rightarrow l^{+} \nu )$, where $Br^{exp}(M^{+} \rightarrow l^{+} \nu )$ and $ Br^{SM}(M^{+} \rightarrow l^{+} \nu )$ express the experimental measurement and SM prediction values of the corresponding branching ratio, respectively, which are taken from Ref. [3]. One can see from Table 1 that the sterile neutrino $N$ can indeed produce correction effects on the branching ratio $ Br(M^{+} \rightarrow l^{+} \nu ) $, and generate large contributions to some specific processes. For example, for the decays $ B^{+} \rightarrow e^{+} \nu_{e} $ and $ D_{s}^{+} \rightarrow e^{+} \nu_{e} $, the values of the ratio $R$ can reach 5.4 and 1.22 for  $ m_{N}=3.5 $ GeV and  1.3 GeV, respectively. However, the contributions of $N$ to most of these decay processes are much smaller than the corresponding experimental uncertainty, or it can not make the value of the branching ratio $ Br(M^{+} \rightarrow l^{+}\nu) $ reach the experimental measurement value, which can not drive new constraints on the scenario with only one sterile neutrino and negligible mixing with the active neutrinos. The exception is  $ Br (K^{+} \rightarrow \mu^{+} \nu_{\mu}) $, which can exceed the corresponding experimental measurement value, thus might give new constraints on the free parameters $ \alpha's $ and $ m_{N} $, as shown in Fig. 1. For 0.12 GeV $ \leq m_{N} \leq $ 0.36 GeV, the sterile neutrino $N$ can make the value of $ Br(K^{+} \rightarrow \mu^{+}\nu_{\mu}) $ exceed the experimental upper limit.

\vspace{0.5cm} \noindent{\bf 2.3 The pseudoscalar meson decays $ M \rightarrow l\nu\gamma$  }

\vspace{0.5cm} Contrary to the pure leptonic decays $ M \rightarrow l\nu $, the radiative leptonic decays $ M \rightarrow l\nu\gamma $ are not subject to the helicity suppression due to the presence of a radiative photon, which might be comparable or even larger than the corresponding decay $ M \rightarrow l\nu $. In the SM, the decay width of $ M^{+}\rightarrow l^{+}\nu\gamma $ with $ l=e $ and $ \mu $ can be general given  at tree-level by [19]
\begin{eqnarray} \Gamma(M^{+} \rightarrow l^{+} \nu\gamma) = \frac{\alpha G_{F}^{2} \left | V_{ij} \right |^{2}}{2592 \pi^{2}} F_{M}^{2} m_{M}^{3} (x_{i}+x_{j})
\end{eqnarray}
with $ x_{i}=(3 -m_{M} / m_{q_{i}})^{2} $  and  $ x_{j}=(3 - 2  m_{M} / m_{q_{j}})^{2} $. Since there are IR divergences in these decay processes when the photon is
soft or collinear with the emitted lepton, the decay widths depend on the experimental resolution to the photon energy. By using the lifetimes of the relevant
mesons, one can obtain the branching ratios for considered decay channels. If one only considers the relative correction effects of new physics on the  decays $ M \rightarrow l\nu\gamma $, then the theoretical uncertainties can be canceled to a large extent. The values of the relative correction parameter $ R'(M^{+} \rightarrow l^{+} \nu \gamma)= Br^{SM+NP}(M^{+} \rightarrow l^{+} \nu \gamma) / Br^{SM}(M^{+} \rightarrow l^{+} \nu \gamma) $ are almost independent of the resolution of the photon energy.

It is well known that the measurement of pure leptonic decays of mesons is very difficult because of the helicity suppression and only one detected final state particle. Since the radiative leptonic decays having an extra real photon emitted in the final state, the reconstruction of these decays is easier to do. Furthermore, the radiative leptonic decays of mesons may be separated properly and compared with measurements directly as long as the theoretical softness of the photon corresponds to the experimental resolutions. In recent years, the experimental studies for the radiative leptonic decays $ M \rightarrow l\nu\gamma $ have been improved greatly, the experimental upper limits for some decay channels are obtained [3]. Certainly, these results depend on the photon threshold energy.

In the scenario considered in this paper, the sterile neutrino $N$ can decay via two kinds of decay channels for $ m_N < m_W$, which are the three-fermion and photon-neutrino
channels. Its decay length can be translated from the total width, which depends on its mass and couplings. If the decay length of the sterile neutrino $N$ is smaller than or  is of the same order of the size of the detector, then it can decay inside the detectors after traveling  a macroscopical distance, its possible signals might be detected via taking advantage of displaced vertex techniques. Although the decay channel $ N \rightarrow \nu\gamma $ is induced by the effective tensorial operators generated at loop level as shown in Eq.(3), it is the dominant decay mode of the sterile neutrino for low $  m_{N} $ and there is $ Br (N \rightarrow \nu\gamma)\approx 1 $ for $ m_{N} < 10 $ GeV [13]. Thus the sterile neutrino $N$ can contribute to the decays $ M \rightarrow l\nu\gamma $ via the process $ M \rightarrow lN \rightarrow l \nu \gamma $. In this subsection, we will calculate its contributions  to the decay processes $ M^{+} \rightarrow l^{+} \nu \gamma $ with $M$ and $l$ denoting $B$, $D$ or $K$ and $e$ or $\mu$ , respectively.

 In the case of neglecting the interference effects between the sterile and active neutrinos, using Eq.(7) we can obtain the contributions of $N$ to the decays $ M^{+} \rightarrow l^{+} \nu \gamma $. Our numerical results show that all of the values of the relative correction parameter $ R'(M^{+} \rightarrow l^{+} \nu \gamma)= Br(M^{+} \rightarrow l^{+} \nu \gamma) / Br^{SM}(M^{+} \rightarrow l^{+} \nu \gamma) $ are smaller than one in a thousand for $ l^{+}=e^{+} $ and $ M^{+}=K^{+}$, $D^{+} $ and $ B^{+} $, while their values can be significantly large for $ l^{+} = \mu^{+}$. This is because we have taken the coupling constants involving the first generation leptons equaling to $ 3.2 \times 10^{-2} \sqrt{m_{N} / 100} $ and other ones equaling to 0.32 for $ \Lambda=1 $ TeV. Our numerical results for $M^{+} \rightarrow \mu^{+} \nu_{\mu} \gamma$ are shown in Fig. 2 and Fig. 3. One can see from these figures that the maximal values of the parameter $ R'$ are 1.08, 2.72 and 5.87 for $M$ being $B$, $D$ and $K$, respectively. Up to now, there is not the experimental measurement value of the branching ratio $ Br (D^{+} \rightarrow \mu^{+} \nu_{\mu} \gamma) $, while the experimental uncertainties for the branching ratio $ Br (K^{+} \rightarrow \mu^{+} \nu_{\mu} \gamma) $ are very large.  The experimental upper limit for the decay $B^{+} \rightarrow \mu^{+} \nu_{\mu} \gamma $ is $3.4\times 10^{-6}$ at $90\%$ CL [3], the value of the parameter $ R'^{exp}$ is smaller than 2.124. We hope that the theoretical calculations and experimental measurements about the radiative leptonic decays of the pseudoscalar mesons $ M \rightarrow l\nu\gamma $ will be improved greatly in near future and the correction effects of the sterile neutrino might be detected in the future $ e^{+} e^{-}$ colliders.

\vspace{0.5cm} \noindent{\bf 3. The sterile neutrino $N$ and the decays $ M \rightarrow \nu\overline{\nu} $ }

\vspace{0.5cm}In the SM the decays $ M \rightarrow \nu\overline{\nu} $ proceed through $Z-penguin$ and electroweak box diagrams. The effective Hamiltonian is [20]
\begin{eqnarray}
H_{eff}= \frac{4 G_{F}}{\sqrt{2}} \frac{\alpha}{2 \pi s_{W}^{2}} \sum_{i=e,\mu,\tau} \sum_{k} \lambda_{k} X^{i}(x_{k})(\overline{q_{L}} \gamma^{\mu} q'_{L}) (\overline{\nu_{L}^{i}} \gamma_{\mu} \nu_{L}^{i}),
\end{eqnarray}
where the functions $ \lambda_{k} X^{i}(x_{k}) $ are relevant combinations of the CKM factors and Inami-Lim functions [21], which depend on the kinds of quarks constituting mesons.

If we take the same assumption as that of subsection 2.2 for the sterile neutrino $N$, then it can contribute to the decays $ M \rightarrow \nu\overline{\nu} $ via the decay  processes $ M \rightarrow \nu_{L}N_{R} $ and $ M \rightarrow N\overline{N} $. In the case of neglecting the sterile-active neutrinos mixing, the $ Z \nu_{L} N_{R} $ coupling can only be induced at loop level, thus the contributions of $ M \rightarrow \nu_{L}N_{R} $ are much smaller than those for the process $ M \rightarrow N\overline{N} $, which can be safely ignored. The leading order Feynman diagrams for the quark level process $ q_{i} \rightarrow q_{j}N\overline{N} $ are shown in Fig. 4, where $ q_{i}= b$, $ c $ and $s$ quarks for the $B$, $D$ and $K$ mesons, respectively.

Using Eq.(2), the expression for the branching ratio $ Br(B_{q} \rightarrow N\overline{N}) $ with $ q= s $ or $ d $ quark can be written as
\begin{eqnarray}
Br(B_{q} \rightarrow N\overline{N})& = & \frac{G_{F}^{2} \alpha^{2} \tau_{Bq}}{8 \pi^{3} s_{W}^{4}} F_{B_{q}}^{2} m_{B_{q}} m_{N}^{2} \sqrt{1 - \frac{4 m_{N}^{2}}{m_{B_{q}}^{2}}}\\
&&\left. \times \{ \lambda_{t} [ \frac{v^{2} \alpha_{Z}}{\Lambda^{2}} C_{0}(x_{t}) - \frac{(\alpha_{W}^{i})^{2} v^{4}}{\Lambda^{4}} B_{0}(x_{t})]+ \lambda_{c} x_{c}^{i}(\frac{\alpha_{W}^{i} v^{2}}{2 \Lambda^{2}})^{2}\}^{2} \right. \nonumber
\end{eqnarray}
with
\begin{eqnarray}C_{0}(x_{t})= \frac{x_{t}}{8} [\frac{3 x_{t}+2}{(x_{t}-1)^{2}}\ln x_{t} + \frac{x_{t}-6}{x_{t}-1}], \end{eqnarray}
\begin{eqnarray}B_{0} (x_{t})=\frac{x_{t}}{4} [\frac{\ln x_{t}}{(x_{t}-1)^{2}} - \frac{1}{x_{t}-1}].    \end{eqnarray}
Where $ \tau_{B_{q}} $ is the lifetime of the pseudoscalar meson $ B_{q} $, $ x_{t}=m_{t}^{2} / m_{W}^{2} $. The CKM combinatorial factors $ \lambda_{t} $ and $ \lambda_{c} $ are $  V_{ts}^{*}  V_{tb} $ and  $  V_{cb}^{*}  V_{cs}  $, $ V_{td}^{*}  V_{tb}  $ and $  V_{cb}^{*}  V_{cd}  $ for the mesons $ B_{s} $ and $ B_{d} $, respectively. In Eq.(11), we have considered the contributions of the box diagrams with the propagating charm quark. The contributions of the $Z-penguin$ diagrams involving the light quarks are neglected. The individual values of $ x_{c}^{i} $ are obtained from Table 1 of Ref. [22]: $ x_{c}^{e, \mu}= 11.8 \times 10^{-4} $, $x_{c}^{\tau}=7.63 \times 10^{-4} $. The calculation formula of the branching ratio $ Br(K_{L} \rightarrow N\overline{N}) $ can be easily given from Eq.(10) via replacing $ B_{q} \rightarrow K_{L} $ and $ \lambda_{t}= V_{ts}^{*} V_{td} $ , $ \lambda_{c}= V_{cs}^{*} V_{cd}  $.

Unlike the decay processes $ B_{q} \rightarrow N\overline{N} $ and $ K_{L} \rightarrow N\overline{N} $, which are dominated by top quark contributions, the decay $ D \rightarrow N\overline{N} $ is mainly induced by the bottom and strange quarks [20, 21]. The expression of the branching  ratio $ Br(D \rightarrow N\overline{N}) $ are

\begin{eqnarray}Br(D \rightarrow N\overline{N})&=&\frac{\alpha^{2} G_{F}^{2} \tau_{D}}{8 \pi^{3} s_{W}^{4}} m_{D} m_{N}^{2} F_{D}^{2} \sqrt{1- \frac{4 m_{N}^{2}}{m_{D}^{2}}} \{ \lambda_{b} [ \frac{\alpha_{Z} v^{2}}{\Lambda^{2}} C(x_{b})\\
&&\left. + \frac{(\alpha_{W}^{i})^{2}  v^{4}}{\Lambda^{4}} B(x_{b},y_{i})]+ \lambda_{s}[ \frac{\alpha_{Z} v^{2}}{\Lambda^{2}}C(x_{s}) + \frac{(\alpha_{W}^{i})^{2} v^{4} }{\Lambda^{4}}B(x_{s},y_{i})]\}^{2} \right.\nonumber  \end{eqnarray}
with
\begin{eqnarray}C(x_{q})=\frac{x_{q}(x_{q}-3)}{4(x_{q}-1)} + \frac{3x_{q}+2}{4(x_{q}-1)^{2}}  x_{q}  \ln x_{q}, \end{eqnarray}
\begin{eqnarray}B(x_{q},y_{i})=-\frac{1}{8} (\frac{y_{i}-4}{y_{i}-1})^{2} x_{q} \ln y_{i} +\frac{x_{q}y_{i} - 8y_{i} + 16}{8(x_{q}-1)^{2}(y_{i}-x_{q})} x_{q}^{2} \ln x_{q} + \frac{(y_{i}-10)x_{q}}{8(y_{i}-1)(x_{q}-1)}.   \end{eqnarray}
Where $ \lambda_{q}=V_{cq}^{*} V_{uq} $, $ x_{q}= m_{q}^{2} / m_{W}^{2} $ and $ y_{i}= m_{l}^{2} / m_{W}^{2} $ with $i$ being the SM leptons. Although it is possible for all of leptons ($ e $, $ \mu $ and $ \tau $) appearing in the box diagrams, there is of numerical significance only when considering the lepton $ \tau $.

It is well known that the branching ratio $ Br(M \rightarrow \nu\overline{\nu}) $ is zero in the SM. Any nonzero measurement of $ Br( M \rightarrow \nu\overline{\nu}) $ would be a clean signal of new physics beyond the SM. If the light sterile neutrino $N$ can not further decay to other particles in the detector , its possible signals may be detected via the decay $ M \rightarrow N\overline{N} $. In the scenario considered in this paper, the branching ratios $ Br( B_{q} \rightarrow N\overline{N}) $, $ Br(K_{L} \rightarrow N\overline{N}) $ and $ Br( D \rightarrow N\overline{N}) $ are plotted as functions of the sterile neutrino mass $ m_{N} $ in Fig. 5, Fig. 6 and Fig. 7, respectively. In our numerical calculation, we have taken $ \Lambda=1 $ TeV,  $ \alpha_{W}^{e}= 3.2 \times 10^{-2} \sqrt{(m_{N}/100)}$ and $ \alpha_{W}^{i}\simeq  \alpha_{Z}=  0.32 $ with $i= \mu$ and $\tau$. From these figures, one can see that the maximal value of $ Br(K_{L} \rightarrow N\overline{N}) $ is $3.5 \times 10^{-13}$, which is still smaller than that given by Ref. [9] in the minimal seesaw models with only one sterile neutrino $N$ and non-negligible active-sterile mixing. For the decay process $ D \rightarrow N\overline{N} $, its value is at the order of $10^{-18}$, which is very difficult to be detected in near future. The branching ratio $ Br( B_{s} \rightarrow N\overline{N}) $ is larger than $ Br( B_{d} \rightarrow N\overline{N}) $ by about two orders of magnitude. Its value can reach $4.2 \times 10^{-10}$, which might approach the sensitivity of Belle II.

\vspace{0.5cm} \noindent{\bf 4. Conclusions }

\vspace{0.5cm}The pure and radiative leptonic decays of the pseudoscalar meson $M$ are theoretically very clean. The only non-perturbation quantity involved in these decay processes is the meson decay constant $ F_{M} $. The decays $ M \rightarrow l\nu $ and $ M \rightarrow \nu\overline{\nu} $ are helicity suppressed in the SM, and the branching ratio $ Br(M \rightarrow \nu\overline{\nu}) $ is exactly zero with massless neutrinos. The radiative  leptonic decays $ M\rightarrow l\nu\gamma $ are not subject to the helicity suppression, which might be comparable or even larger than the corresponding decay $ M \rightarrow l\nu $. Thus, all of these processes are sensitive to new physics effects. It is very interest to study the contributions of new physics to these decay processes and see whether can be detected in future collider experiments.

Many new physics models giving the tiny neutrino masses predict the existence of the sterile neutrinos with mass covering various mass ranges.  Recently, the relatively light sterile neutrinos with masses at the GeV scale have been attracting some interests. In this paper we consider a scenario with only one sterile neutrino  $N $ of negligible mixing with the active neutrinos, where the sterile neutrino interactions could be described in a model independent approach based on an effective theory. Under such a framework, we consider the contributions of the sterile neutrino $N$ to the decays $ M \rightarrow l\nu$, $ l\nu\gamma $  and $ \nu\overline{\nu} $ with $ M$ being the pseudoscalar mesons $B$, $D$ and $K$. Our numerical results show:

1. The sterile neutrino $N$ can indeed enhance the values of the branching ratios $ Br (M \rightarrow l\nu) $ predicted by the SM. However, for most of these decay processes, it can not make $ Br(M \rightarrow l\nu) $ reach the experimental measurement value. The exception is  $ Br (K^{+} \rightarrow \mu^{+} \nu_{\mu}) $, which can exceed the corresponding current experimental up limit, thus might give new constraints on the free parameters $ \alpha's $ and $ m_{N} $.

2. The contributions of the sterile neutrino $N$ to the decays $ M^{+} \rightarrow e^{+}\nu_{e}\gamma $ with $M^{+}= B^{+}$, $K^{+} $ and $D^{+}$ are very small and the values of the relative correction parameter $R'(M^{+} \rightarrow e^{+}\nu_{e}\gamma)$ are smaller than  one in a thousand. While it can produce significant contributions to $ M^{+} \rightarrow \mu^{+}\nu_{\mu}\gamma $, which can make the values of the parameters $R'(B^{+} \rightarrow \mu^{+}\nu_{\mu}\gamma)$, $R'(K^{+} \rightarrow \mu^{+}\nu_{\mu}\gamma)$ and  $R'(D^{+} \rightarrow \mu^{+}\nu_{\mu}\gamma)$ reach 1.08, 5.87 and 2.72, respectively.

3. All of the branching ratios $ Br(M \rightarrow \nu\overline{\nu})$ with $ M$ being the pseudoscalar mesons $B$, $D$ and $K$ can be significant enhanced by the sterile neutrino $N$. For the branching ratio $ Br( B_{s} \rightarrow \nu\overline{\nu})$, its value can reach $4.2 \times 10^{-10}$, which might approach the sensitivity of Belle II.

\section*{Acknowledgments} \hspace{5mm}This work was supported in part by the National Natural Science Foundation of China under Grant No. 11275088
and 11875157.

\begin{table}[hp]
\renewcommand{\arraystretch}{0.7}
\begin{center}
\begin{tabular}{|c|c|c|c|c|c|}
\hline
$meson$ & $mode$ & $m_{N}(GeV)$ & $R$ & \multicolumn{2}{|c|}{$R^{exp}$ }  \\
\cline{5-6}
\multirow{2}{*}{} & \multirow{2}{*}{} & \multirow{2}{*}{} & \multirow{2}{*}{} & $min$ & $max$   \\
\hline
\multirow{6}{*}{$ K^{+} $} & \multirow{3}{*}{} &  0.34 & 1.00353 & \multirow{3}{*}{} & \multirow{3}{*}{}  \\
\cline{3-4}
\multirow{6}{*}{} & $   e^{+}\nu_{e}  $ &  0.38 & 1.00297 & $ 1.00191$ & $1.01081$ \\ \cline{3-4}
\multirow{6}{*}{} &  \multirow{3}{*}{} & 0.42 & 1.00184 &  \multirow{3}{*}{} &   \multirow{3}{*}{}  \\
\cline{2-6}
\multirow{6}{*}{} & \multirow{3}{*}{} &  0.30 & 1.0033 & \multirow{3}{*}{} & \multirow{3}{*}{}  \\
\cline{3-4}
\multirow{6}{*}{} & $ \mu^{+}\nu_{\mu} $ &  0.34 & 1.00259 & $ 0.998426$ & $1.00189$ \\ \cline{3-4}
\multirow{6}{*}{} &  \multirow{3}{*}{} & 0.38 & 1.00103 &  \multirow{3}{*}{} &   \multirow{3}{*}{}  \\
\hline
\multirow{9}{*}{$ D^{+} $} & \multirow{3}{*}{} &  1.2 & 1.19442 & \multirow{3}{*}{} & \multirow{3}{*}{}  \\
\cline{3-4}
\multirow{9}{*}{} & $ e^{+}\nu_{e} $ &  1.4 & 1.17226& $ {-}$ & $ 1517.24$ \\ \cline{3-4}
\multirow{9}{*}{} &  \multirow{3}{*}{} & 1.6 & 1.0954 &  \multirow{3}{*}{} &   \multirow{3}{*}{}  \\
\cline{2-6}
\multirow{9}{*}{} & \multirow{3}{*}{} &  1.2 & 1.03839 & \multirow{3}{*}{} & \multirow{3}{*}{}  \\
\cline{3-4}
\multirow{9}{*}{} & $ \mu^{+}\nu_{\mu} $ &  1.4 & 1.02887& $ 1.44534$ & $ 1.583$ \\ \cline{3-4}
\multirow{9}{*}{} &  \multirow{3}{*}{} & 1.6 & 1.01337 &  \multirow{3}{*}{} &   \multirow{3}{*}{}  \\
\cline{2-6}
\multirow{9}{*}{} & \multirow{3}{*}{} &  0.02 & 1.00083 & \multirow{3}{*}{} & \multirow{3}{*}{}  \\
\cline{3-4}
\multirow{9}{*}{} & $ \tau^{+}\nu_{\tau} $ &  0.04 & 1.00077& $ {-}$ & $  1.39535$ \\ \cline{3-4}
\multirow{9}{*}{} &  \multirow{3}{*}{} & 0.06 & 1.00067 &  \multirow{3}{*}{} &   \multirow{3}{*}{}  \\
\hline
\multirow{9}{*}{$ D_{s}^{+} $} & \multirow{3}{*}{} &  1.0 & 1.1792 & \multirow{3}{*}{} & \multirow{3}{*}{}  \\
\cline{3-4}
\multirow{9}{*}{} & $ e^{+}\nu_{e} $ &  1.3 & 1.2275& $ {-}$ & $882.979$ \\ \cline{3-4}
\multirow{9}{*}{} &  \multirow{3}{*}{} & 1.6 & 1.15381 &  \multirow{3}{*}{} &   \multirow{3}{*}{}  \\
\cline{2-6}
\multirow{9}{*}{} & \multirow{3}{*}{} &  1.0 & 1.04267 & \multirow{3}{*}{} & \multirow{3}{*}{}  \\
\cline{3-4}
\multirow{9}{*}{} & $ \mu^{+}\nu_{\mu} $ &  1.3 & 1.04136& $ 1.3175$ & $ 1.4325$ \\ \cline{3-4}
\multirow{9}{*}{} &  \multirow{3}{*}{} & 1.6 & 1.02222 &  \multirow{3}{*}{} &   \multirow{3}{*}{}  \\
\cline{2-6}
\multirow{9}{*}{} & \multirow{3}{*}{} &  0.10 & 1.00075 & \multirow{3}{*}{} & \multirow{3}{*}{}  \\
\cline{3-4}
\multirow{9}{*}{} & $ \tau^{+}\nu_{\tau} $ &  0.13 & 1.00067& $ 1.38889$ & $  1.51058$ \\ \cline{3-4}
\multirow{9}{*}{} &  \multirow{3}{*}{} & 0.16 & 1.00052 &  \multirow{3}{*}{} &   \multirow{3}{*}{}  \\
\hline
\multirow{9}{*}{$ B^{+} $} & \multirow{3}{*}{} &  3.0 & 5.02811 & \multirow{3}{*}{} & \multirow{3}{*}{}  \\
\cline{3-4}
\multirow{9}{*}{} & $ e^{+}\nu_{e} $ & 3.5 & 5.38293& $ {-}$ & $  113426$ \\ \cline{3-4}
\multirow{9}{*}{} &  \multirow{3}{*}{} & 4.0 & 4.77823 &  \multirow{3}{*}{} &  \multirow{3}{*}{}  \\
\cline{2-6}
\multirow{3}{*}{} & \multirow{3}{*}{} &  3.0 & 1.31367 & \multirow{3}{*}{} & \multirow{3}{*}{}  \\
\cline{3-4}
\multirow{9}{*}{} & $ \mu^{+}\nu_{\mu} $ &  3.5 & 1.29235& ${-}$ & $  2.7027$ \\ \cline{3-4}
\multirow{9}{*}{} &  \multirow{3}{*}{} & 4.0 & 1.22025 &  \multirow{3}{*}{} &   \multirow{3}{*}{}  \\
\cline{2-6}
\multirow{9}{*}{} & \multirow{3}{*}{} & 2.5 & 1.00179 & \multirow{3}{*}{} & \multirow{3}{*}{}  \\
\cline{3-4}
\multirow{9}{*}{} & $ \tau^{+}\nu_{\tau} $ &  3.0 & 1.00154& $ 1.02594$ & $  1.47406$ \\ \cline{3-4}
\multirow{9}{*}{} &  \multirow{3}{*}{} & 3.5 & 1.00012&  \multirow{3}{*}{} &   \multirow{3}{*}{}  \\
\hline
\end{tabular}
\caption{The values of the ratio $R$ induced by the sterile neutrino $N$ for the decay \hspace*{1.78cm} $M^{+} \rightarrow l^{+}\nu $ with different mass $m_{N}$. The fifth and sixth columns express the \hspace*{1.78cm}minimum and maximum values of the parameter $R^{exp}$, respectively.}
\end{center}
\end{table}

\begin{figure}[htb]
\vspace{-0.5cm}
\begin{center}
 \epsfig{file=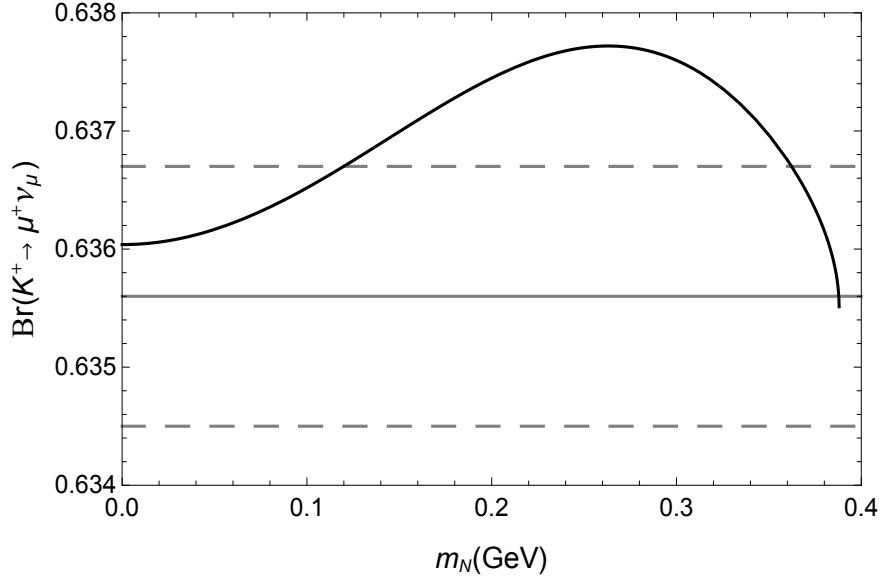, width=4.5in,height=3in}
 \vspace{-0.5cm}
 \caption{The branching ratio $Br(K^{+} \rightarrow \mu^{+} \nu_{\mu} $) as a function of the mass $m_{N}$. The  \hspace*{2cm}region
between horizontal dashed lines correspond $1\sigma$ allowed region from the  \hspace*{2cm}experimental measurement value of $Br(K^{+} \rightarrow \mu^{+} \nu_{\mu} $).}
 \label{ee}
\end{center}
\end{figure}

 \begin{figure}[htb]
 \vspace{-0.5cm}
 \begin{center}
  \epsfig{file=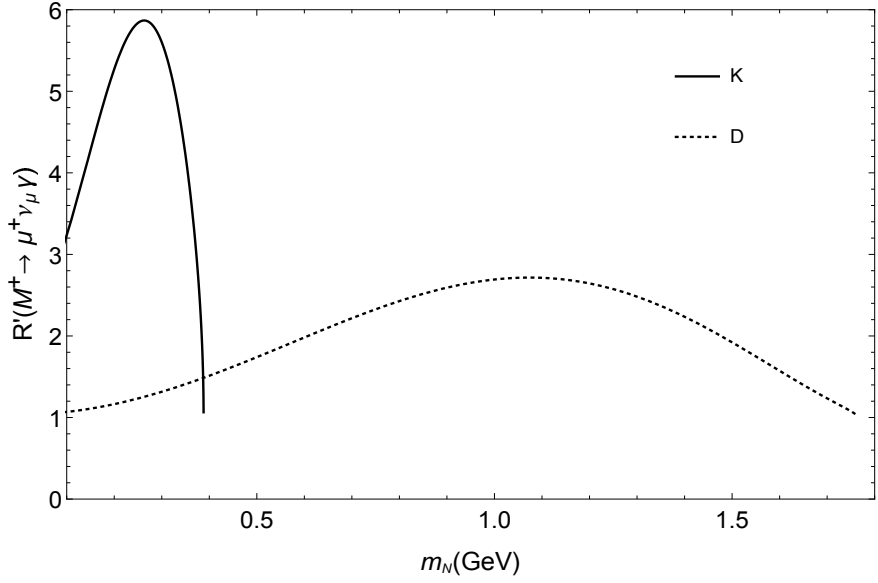, width=4.5in,height=3in}
  \vspace{-0.5cm}
 \caption{The relative correction parameter $R'$  as a function of the mass $m_{N}$  for  the \hspace*{2cm}decay processes
$K^{+}\rightarrow \mu^{+}\nu_{\mu}\gamma$ (solid line) and $D^{+}\rightarrow \mu^{+}\nu_{\mu}\gamma$ (dotted line).}
 \label{ee}
 \end{center}
 \end{figure}

 \begin{figure}[htb]
 \vspace{-0.5cm}
 \begin{center}
  \epsfig{file=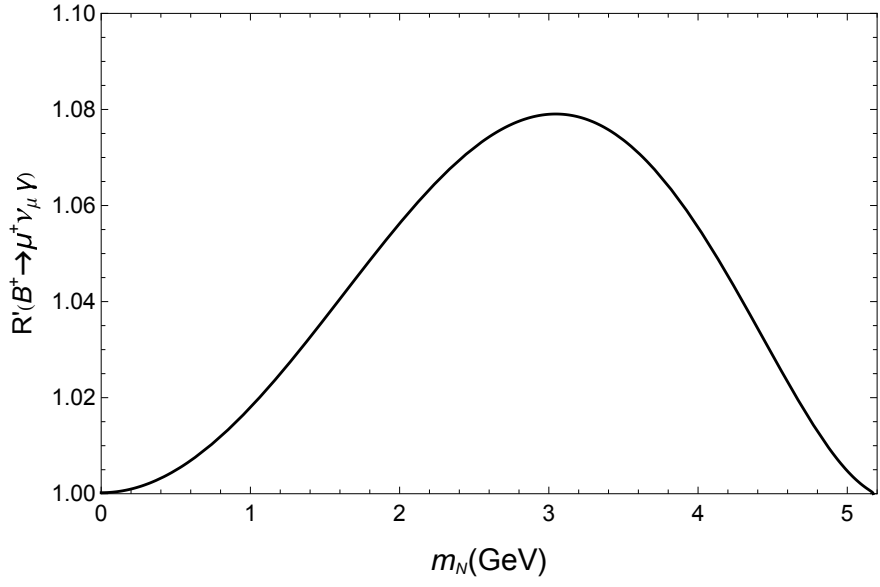, width=4.5in,height=3in}
  \vspace{-0.5cm}
 \caption{The relative correction parameter $R'$  as a function of the mass $m_{N}$ for \hspace*{2cm}the radiative decay process
 $B^{+}\rightarrow \mu^{+}\nu_{\mu}\gamma$.}
 \label{ee}
 \end{center}
 \end{figure}

\begin{figure}[h]
\begin{tabular}{cc}
\begin{minipage}[h]{3in}
\includegraphics[width=3in]{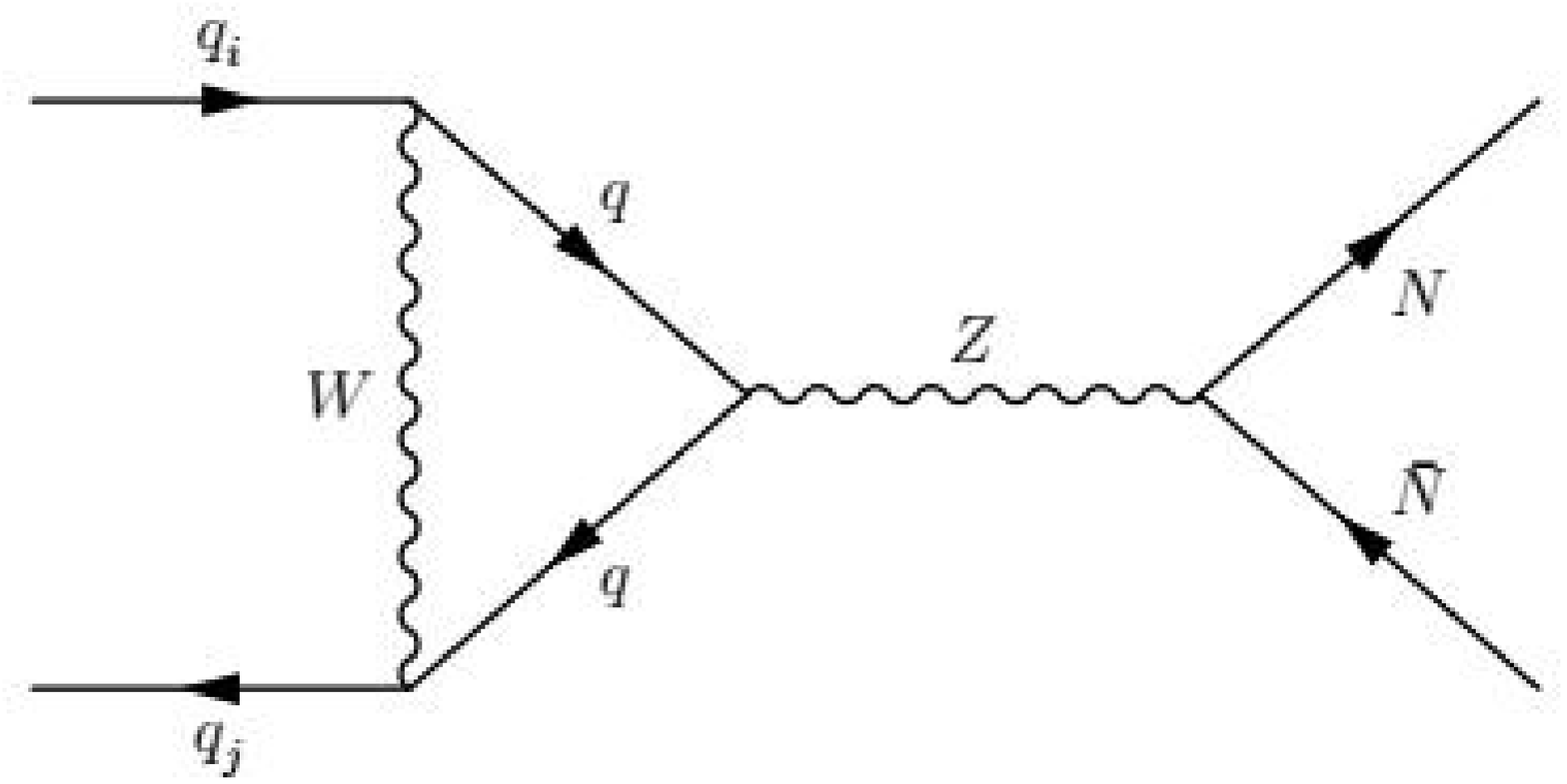}
\end{minipage}
\begin{minipage}[h]{3in}
\includegraphics[width=3in]{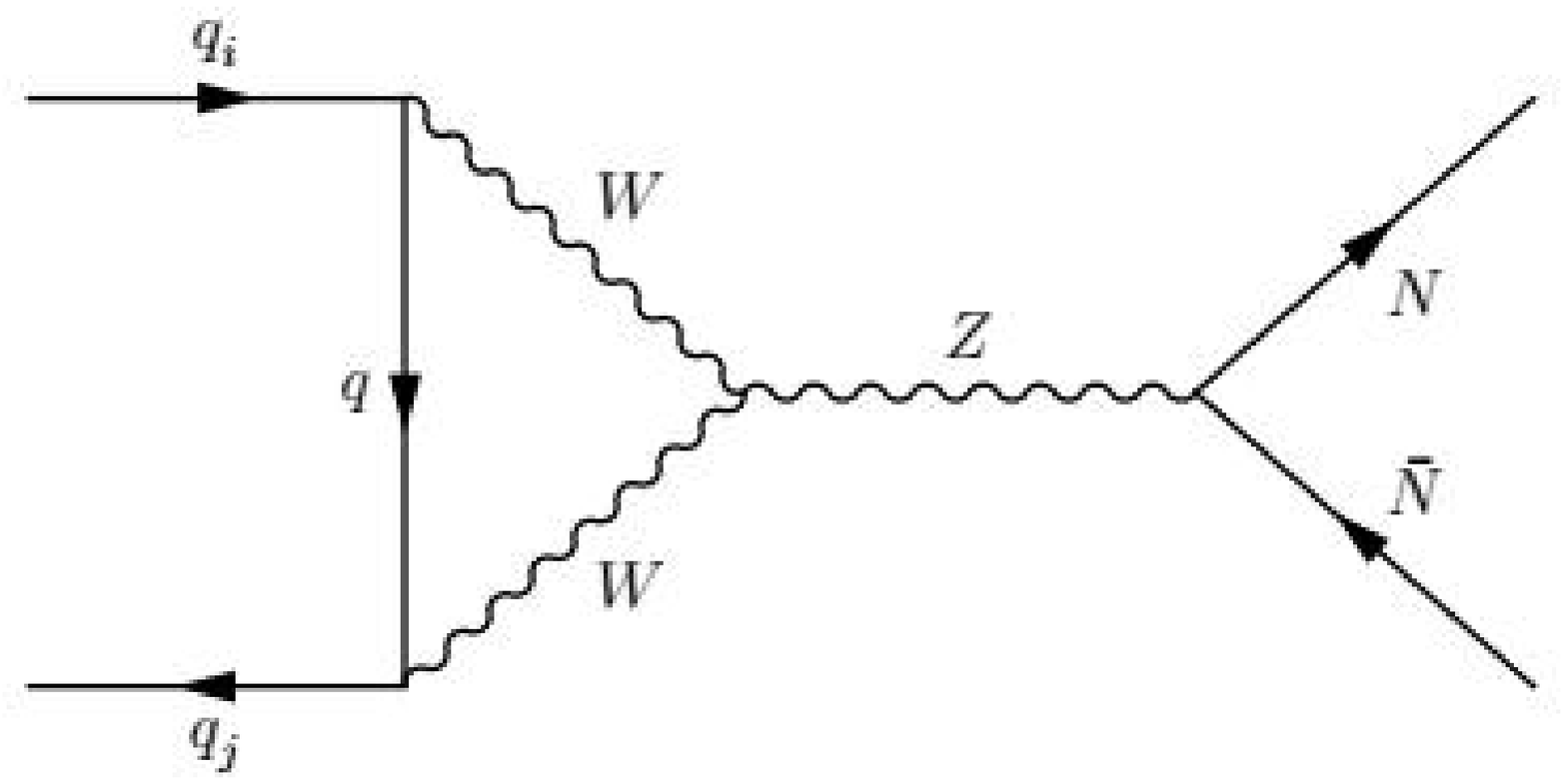}
\end{minipage} \\
\begin{minipage}[h]{3in}
\includegraphics[width=3in]{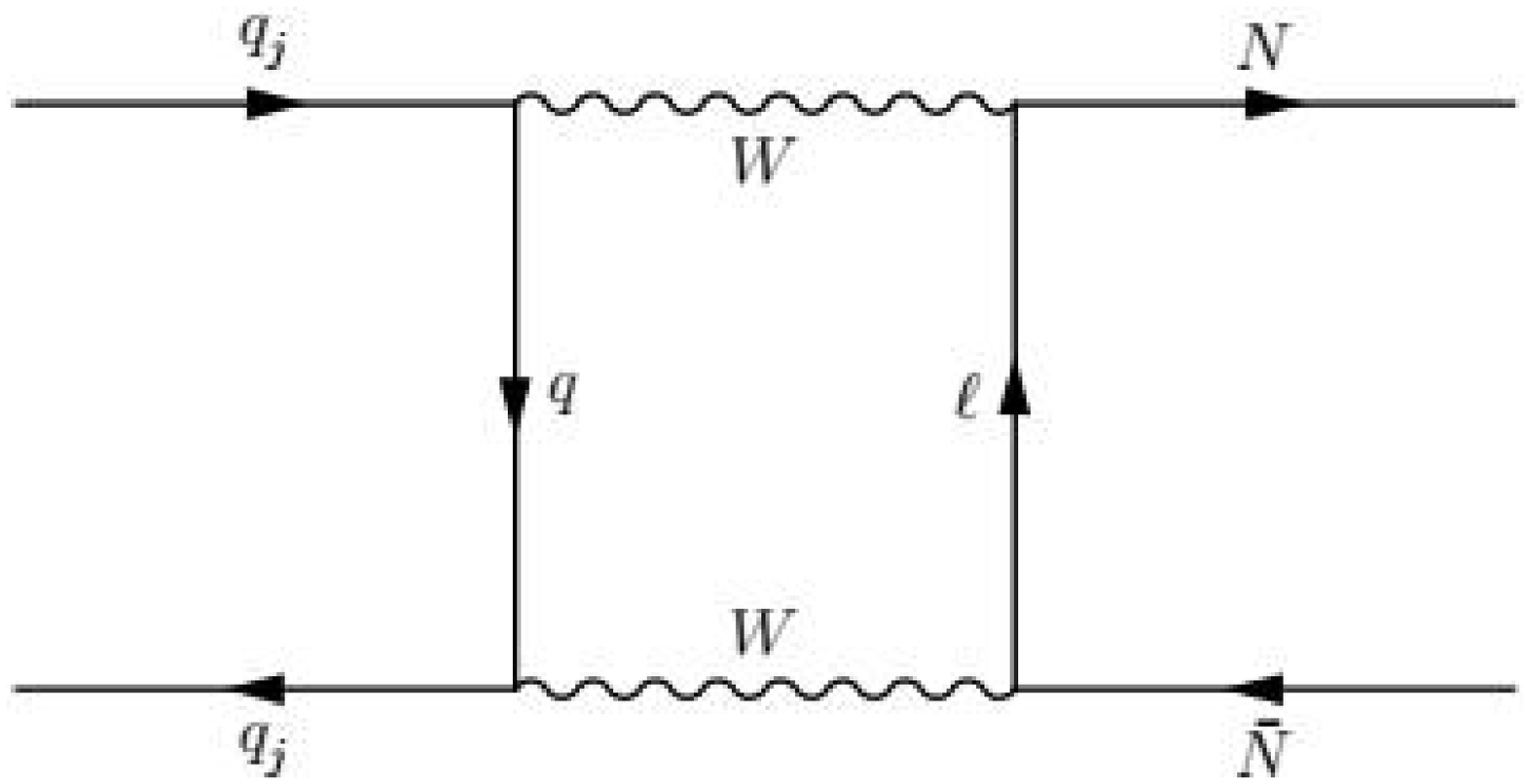}
\end{minipage}
\end{tabular}
 \caption{Leading order Feynman diagrams for the process $ q_{i} \rightarrow q_{j} N\overline{N} $}
\end{figure}

 \begin{figure}[htb]
 \vspace{-0.5cm}
 \begin{center}
  \epsfig{file=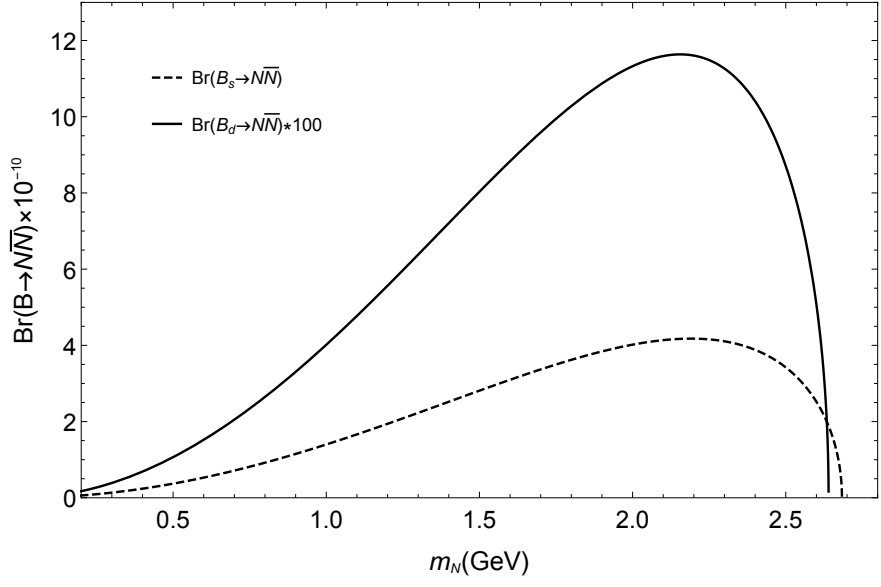, width=4.5in,height=3in}
  \vspace{-0.5cm}
 \caption{The branching ratio $Br(B_{q} \rightarrow N\overline{N})$ as a function of the mass $m_{N}$ for \hspace*{2cm} $ q=s $ (solid line)
and $d$ (dashed line) quarks.}
 \label{ee}
 \end{center}
 \end{figure}

\begin{figure}[htb]
 \vspace{-0.5cm}
 \begin{center}
  \epsfig{file=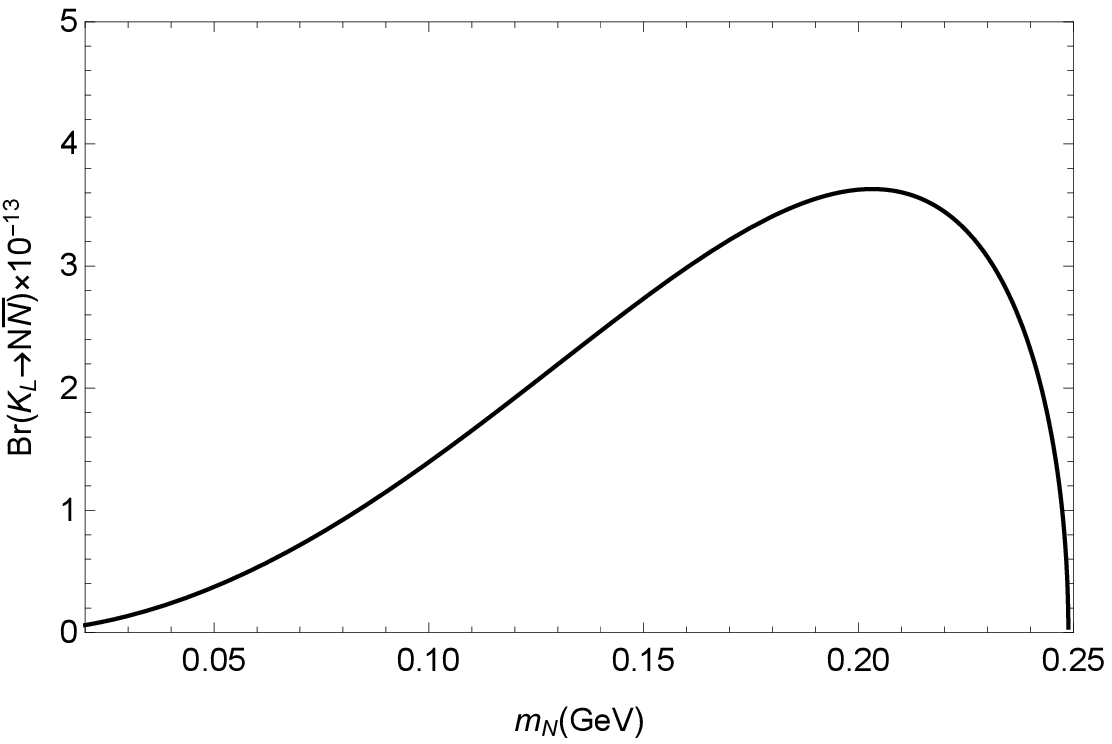, width=4.5in,height=3in}
  \vspace{-0.5cm}
 \caption{The branching ratio $Br(K_{L} \rightarrow N\overline{N})$ as a function of the mass $m_{N}$.}
 \label{ee}
 \end{center}
 \end{figure}

 \begin{figure}[htb]
  \vspace{-0.5cm}
  \begin{center}
   \epsfig{file=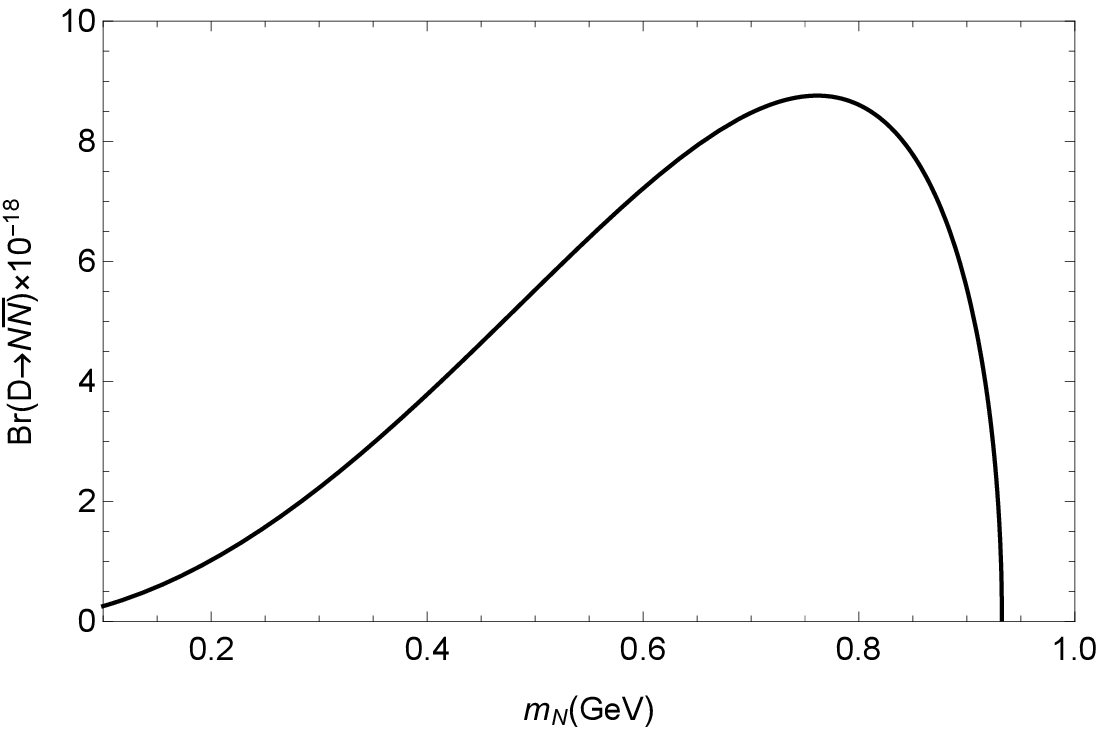, width=4.5in,height=3in}
   \vspace{-0.5cm}
  \caption{The branching ratio $Br(D \rightarrow N\overline{N})$ as a function of the mass $m_{N}$.}
  \label{ee}
  \end{center}
  \end{figure}


\vspace{5mm}


\begin{thebibliography}{99}
\bibitem{1}
S. Chatrchyan, et al. [CMS Collaboration], Phys. Lett. B716, 30 (2012).
\bibitem{2}
G. Aad, et al. [ATLAS Collaboration], Phys. Lett. B716, 1 (2012).
\bibitem{3}
C. Patrignani, et al. [Particle Data Group], Chin. Phys. C40, 100001 (2016), and updated version.
\bibitem{4}
P. Minkowski, Phys. Lett. B67, 421 (1977); R. N. Mohapatra and G. Senjanovic, Phys. Rev. Lett. 44, 912 (1980); J. Schechter and J. W. F. Valle, Phys. Rev. D22, 2227 (1980); K. N. Abazajian, et al., arXiv: 1204.5379 [hep-ph].
\bibitem{5}
 M. Mitra, R. Ruiz, D. J. Scott and M. Spannowsky,  Phys. Rev. D94, 095016 (2016); L. Duarte, J. Peressutti, O. A. Sampayo, J. Phys. G45, 025001 (2018); E. Accomando et al., JHEP 1704, 081 (2017);  P. S. B. Dev, R. N. Mohapatra and Y. Zhang, Nucl. Phys. B923, 179 (2017); S. Antusch, E. Cazzato, and O. Fischer, Phys. Lett. B774, 114 (2017); S. Dube, D. Gadkari, and A. M. Thalapillil, Phys. Rev. D96, 055031 (2017);  E. Accomando, L. Delle Rose, S. Moretti, E. Olaiya and C. H. Shepherd-Themistocleous, JHEP 1802, 109 (2018); G. Cottin, J. C. Helo, M. Hirsch,  Phys. Rev. D97, 055025(2018); C. O. Dib, C. S. Kim, N.  A. Neill, Xing-Bo Yuan, Phys. Rev. D97, 035022 (2018); J. C. Helo, M. Hirsch, Z. S. Wang, JHEP 1807, 056 (2018); S. Jana, N. Okada, D. Raut,  arXiv:1804.06828 [hep-ph]; G. Cottin, J. C. Helo, Martin. Hirsch, Phys. Rev. D98, 035012 (2018) ; A. Abada, N. Bernal, M. Losada, X. Marcano, arXiv:1807.10024 [hep-ph]; Xabier Marcano, arXiv:1808.04705 [hep-ph].
\bibitem{6}
J. C. Helo, M. Hirsch, S. Kovalenko, Phys. Rev. D89, 073005 (2014); E. Izaguirre and B. Shuve, Phys. Rev. D91, 093010 (2015); A. M. Gago, P. Hern$\acute{a}$ndez, J. Jones-P$\acute{e}$rez,  M. Losada, and A. Moreno Brice$\tilde{n}$o, Eur. Phys. J. C75, 470 (2015); S. Antusch, E. Cazzato, O. Fischer, JHEP 1612, 007 (2016); 
 S. Antusch, E. Cazzato, and O. Fischer, Int. J. Mod. Phys. A32, 1750078 (2017); A. Das, N. Okada and D. Raut,  Phys. Rev. D97, 115023 (2018); A. Das, P. S. B. Dev and R. N. Mohapatra, Phys. Rev.D97, 015018 (2018).
 \bibitem{7}
 R. E. Shrock, Phys Lett. B 96, 159 (1980); Phys. Rev. D24, 1232 (1981); R. E. Shrock and M. Suzuki, Phys. Lett. B 112, 382 (1982).
\bibitem{8}
C. Dib and C. S. Kim, Phys. Rev. D89, 077301 (2014); G. Cveti$\check{c}$, C. S. Kim, Phys. Rev. D94, 053001 (2016); G. Cveti$\check{c}$, C. S. Kim, Phys. Rev. D96, 035025 (2017); G. Cveti$\check{c}$, F. Halzen, C. S. Kim, S. Oh, Chin. Phys. C41, 113102 (2017); J. Mejia-Guisao, D. Milan$\acute{e}$s, N. Quintero, J. D. Ruiz-$\acute{A}$lvarez, Phys. Rev. D96, 015039 (2017); J. Mejia-Guisao, D. Milanes, N. Quintero, J. D. Ruiz-$\acute{A}$lvarez, Phys. Rev. D96, 015039 (2017); Han Yuan, Tian-Hong Wang, Yue Jiang, Qiang Li, Guo-Li Wang, J. Phys. G45, 065002 (2018)  ;  A. Abada, V. D. Romeri, M. Lucente, A. M. Teixeira, T. Toma, JHEP 1802, 169 (2018); K. Bondarenko, A. Boyarsky, D. Gorbunov, O. Ruchayskiy, arXiv:1805.08567 [hep-ph]; A. Azatov, D. Barducci, D. Ghosh, D. Marzocca, L. Ubaldi, arXiv:1807.10745 [hep-ph].
\bibitem{9}
 A. Abada, D. Becirevic, O. Sumensari, C. Weiland and R. Z. Funchal, Phys. Rev. D95, 075023 (2017).
\bibitem{10}
F. del Aguila, S. Bar-Shalom, A. Soni and J. Wudka, Phys. Lett. B670, 399 (2009).
\bibitem{11}
S. Bhattacharya and J. Wudka, Phys. Rev. D94, 055022 (2016), [Erratum: Phys. Rev. D95, 039904 (2017)]; P. Ballett, S. Pascoli and M. Ross-Lonergan, JHEP 1704, 102 (2017); Y. Liao and X.-D. Ma,  Phys. Rev. D96, 015012 (2017) ; A. Caputo, P. Hernandez, J. Lopez-Pavon and J. Salvado, JHEP 1706, 112 (2017).
\bibitem{12}
J. Peressutti, I. Romero and O. A. Sampayo, Phys. Rev. D84, 113002 (2011); J. Peressutti and O. A. Sampayo, Phys. Rev. D90, 013003 (2014); L. Duarte, G. A. Gonz$\acute{a}$lez-Sprinberg  and O. A. Sampayo, Phys. Rev. D91, 053007 (2015); L. Duarte, J. Peressutti and O. A. Sampayo, J. Phys. G45, 025001 (2018); C. X. Yue, Y. C. Guo, Z. H. Zhao; Nucl. Phys. B925, 186 (2017); L. Duarte, J. Herrera, G. Zapata, O. A. Sampayo, Eur.Phys.J. C78, 352 (2018).
\bibitem{13}
L. Duarte, J. Peressutti and O. A. Sampayo, Phys. Rev. D92, 093002 (2015); L. Duarte, I. Romero, J. Peressutti and O. A. Sampayo, Eur. Phys. J. C76, 453 (2016).
\bibitem{14}
J. Wudka, AIP Conf. Proc. 531, 81 (2000), hep-ph/0002180.
\bibitem{15}
C. Arzt, M. Einhorn, and J. Wudka, Nucl. Phys. B433, 41 (1995).
\bibitem{16}
D. Silverman, H. Yao, Phys. Rev. D38, 214 (1988).
\bibitem{17}
M. Antonelli et al., Phys. Rept. 494, 197 (2010); V. Cirigliano and I. Rosell, JHEP 0710, 005 (2007); V. Cirigliano, G. Ecker, H. Neufeld, A. Pich, and J. Portoles,  Rev. Mod. Phys. 84, 399 (2012); J. L. Rosner, S. Stone and R. S. Van de Water, Particle Data Book, arXiv: 1509.02220 [hep-ph].
\bibitem{18}
Xiao-Gang He, S. Oh, JHEP 0909, 027 (2009) ; A. Filipuzzi, G. Isidori, Eur. Phys. J. C64, 55 (2009); M. Gomez-Bock, G. Lopez-Castro , L. Lopez-Lozano, A. Rosado, Phys. Rev. D80, 055017 (2009); F. J. Botella et al, JHEP 1407, 078 (2014) .
\bibitem{19}
C. D. Lu, G. L. Song, Phys. Lett. B562, 75 (2003).
\bibitem{20}
For example see: A. Badin,  A. A. Petrov, Phys. Rev. D82, 034005 (2010).
\bibitem{21}
T. Inami and C. S. Lim, Prog. Theor. Phys. 65, 297 (1981) [Erratum-ibid. 65, 1772 (1981)].
\bibitem{22}
G. Buchalla and A. J. Buras,  Nucl. Phys. B412, 106 (1994); R. Allahverdi et al., Phys. Rev. D95, 075001 (2017).

\end{thebibliography}
\end{document}